\documentclass[showpacs,prl,superscriptaddress,nofootinbib, twocolumn]{revtex4}
\usepackage{graphicx}
\usepackage{color}
\usepackage{amssymb,amsfonts,amsmath}
\usepackage{natbib}
\usepackage{epstopdf}
\usepackage{subfigure}

\begin{document}

\title{Reactive explorers to unravel network topology}

\author{Ihusan Adam}  \affiliation{Dipartimento di Ingegneria dell'Informazione, Universit\`{a} di Firenze,
Via S. Marta 3, 50139 Florence, Italy}
\affiliation{Universit\`{a} degli Studi di Firenze, Dipartimento di Fisica e Astronomia,
CSDC and INFN, via G. Sansone 1, 50019 Sesto Fiorentino, Italy}
\author{Duccio Fanelli}  \affiliation{Dipartimento di Fisica e Astronomia and CSDC, Universit\`{a} degli Studi di Firenze, via G. Sansone 1, 50019 Sesto Fiorentino, Italia}
\affiliation{INFN Sezione di Firenze, via G. Sansone 1, 50019 Sesto Fiorentino, Italia}
\author{Timoteo Carletti}
\affiliation{aXys, Namur Institute for Complex Systems, University of Namur, Belgium}
\author{Giacomo Innocenti}
\affiliation{Dipartimento di Ingegneria dell'Informazione, Universit\`{a} di Firenze,
Via S. Marta 3, 50139 Florence, Italy}

\begin{abstract}
A procedure is developed and tested to recover the distribution of connectivity of an a priori unknown network, by sampling the 
dynamics of an ensemble made of reactive walkers. The relative weight between reaction and relocation is gauged by a scalar control parameter, which can be adjusted 
at will. Different equilibria are attained by the system, following the externally imposed modulation, and reflecting the interplay between
reaction and diffusion terms. The information gathered on the observation node is used to predict the
stationary density as displayed by the system, via a direct implementation of the celebrated
Heterogeneous Mean Field (HMF) approximation. This knowledge translates into a linear problem which can be solved to return 
the entries of the sought distribution. A variant of the model is then considered which consists in assuming 
a localized source where the reactive constituents are injected, at a
rate that can be adjusted as a stepwise function of time. The linear problem obtained when operating in this setting 
allows one to recover a fair estimate of the underlying system size. Numerical experiments are carried so as to challenge the predictive 
ability of the theory. 
\end{abstract}

\pacs{} 

\maketitle

\section{Introduction}

Networks are abstract mathematical structures, often invoked in modeling the dynamics of complex interacting units \cite{newman, latora, albert, vespignani, caldarelli}.  The brain, Internet and the cyberworld, foodwebs and social contacts are examples, drawn from distinct fields of investigation, which can be ideally grouped under the unifying umbrella of network science. Nodes (vertices) can point to individual actors of the inspected dynamics (e.g. material units, bits of information, or, on a different scale, extended populations), while edges (links) stand for existing bilateral ties. Alternatively, nodes can tag spatial or functional niches, bounded regions of an embedding landscape, 
 mutually connected by physical or virtual paths, as epitomized by the links~\cite{vespignani,noh,sinatra,nicosia,masuda,latora_mob}. In several cases of interest, punctual entities, also termed agents, may jump from one node to any of its adjacent 
 neighbors, following the intricate network's architecture. Agents relocating across the network via multiple successive jumps are said to execute a random walk: their asymptotic distribution convey important information on the inherent organization of the underlying network. If walkers are transparent to each others, their steady state distribution reflects in fact the degree of connectivity of the nodes, a direct measure of the number of links possessed by any given node of the collection. Photographing the asymptotic nodes' occupancy, enables hence to reconstruct the distribution of connectivities, a topological quantity of paramount importance when aiming at classifying the characteristics of the underlying graph.  Indeed, the structure of the network is often unknown and several methods have been devised in the literature to recover it, from functions back to structure, a non trivial task that hides formidable challenges~\cite{nature,pnas,livi, livi1,livi2,  lucilla1, lucilla2, timme, irene}. Efficient schemes should gather the necessary information from a limited number of nodes, as 
 monitoring the population on each vertex becomes virtually impracticable, for large network sizes. In a recent paper~\cite{prl_crowd} a variant of the random walk problem was introduced which accounts for the mutual interference between agents, as stemming from the competition for the available space in crowded operating condition~\cite{ligget,almaas,kwon,fanelli,GalantiCrowding,Fernando,Landman,Galanti,Galanti2}. Nodes are assigned a finite carrying capacity, a sensible constraint which makes walkers dynamically intertwingled, through dedicated nonlinear terms. The asymptotic density distribution of walkers in the presence of crowding differs significantly from that obtained under diluted conditions. In  crowded conditions, the equilibrium concentration saturates for large enough values of the connectivity. This observation opens up the perspective of recovering the unknown distribution of connectivities from repeated single-node measurements of the asymptotic dynamics, at increasing crowding. The nonlinearities that originate from the interference among microscopic agents competing for space is the key of success to the proposed approach. Building on this achievement, we here generalize the method to the setting where agents perform standard, hence linear, diffusion, but the nonlinearity comes from a local reaction term. In the first part of this paper, the relative strength of the reaction and diffusion contributions is weighted by a scalar control parameter.  Different equilibria are attained
by the system, by modulating the latter parameter, and reflecting the interplay between reaction and diffusion terms. The equilibrium distribution is sampled by punctual measurements performed on
just one node. The information gathered on the observation node is used to predict the
stationary density as displayed by the system, via a direct implementation of the celebrated
Heterogeneous Mean Field (HMF) approximation \cite{vespignani, HMF1, pittorino}. The entries of the sought distribution link the
solution, as obtained within the HMF working ansatz, to the average density sampled on the reference
node. Solving the ensuing linear problem with standard optimization tools, returns a rather accurate
estimate of the distribution of connectivity, as we shall prove for a selected gallery of test network models. In the second part of the paper, we consider a variant of the model by accounting for the presence of  
a source where the reactive constituents are injected, at a rate that we assume to be modulated as a stepwise function of time.  This allows for the fixed point to be successively tweaked, as required by the  reconstruction scheme here developed.  In our application, the reaction
model is assumed of the logistic type and it is therefore tempting to ideally interpret the reactive
explorers, as living entities crawling on the unknown network support.

\section{The mathematical framework}
 
Label with  $x_i$ the concentration of the reactive species on node $i$. The dynamics of the system that we shall examine is governed by the following set of ordinary differential equations: 

\begin{equation}
\label{eq0}
 \dot{x}_i = \alpha f(x_i) +(1-\alpha) \sum_{j=1}^{N}L_{ij}x_j
\end{equation}
\\
where  $L_{ij}=\frac{A_{ij}}{k_j}-\delta_{ij}$ are the entries of the random walk Laplacian operator $\mathbf{L}$; $\mathbf{A}$ is the  adjacency matrix of the (undirected) network, 
while $k_i=\sum_{j}A_{ij}$ denotes the connectivity (or degree) of node $i$. The scalar parameter $\alpha \in [0,1]$ gauges the relative weight of the two terms, appearing on the right hand side of the above equation. In the following,  we will operate under an idealized setting and assume that $\alpha$ can be freely tuned within the interval of pertinence. This choice has pedagogical value, and builds on the analysis in \cite{cencetti}: when  $\alpha=0$ the reaction term is silenced and  the agents behave as standard linear walkers. By making $\alpha$  progressively larger, nonlinearities gain in relevance. The linear problem that is obtained when $\alpha=0$ can be solved analytically. By direct inspection, it is immediate to conclude that, at equilibrium ($\dot{x}_i=0$), $x_i=k_i/\sum_j(k_j)$. When nonlinearities come into play ($\alpha \ne 0$), the complexity of the problem rises considerably and no closed form solutions exist in general. Approximate techniques can be however put forward, to access information on the asymptotic fate of the system. In particular, for relatively small values of $\alpha$ it can be reasonably hypothesized that the displayed concentration is still arranged in classes of connectivities, as it happens in the reference setting of a pure random walk ($\alpha=0$). This working ansatz motivates recasting the problem at hand in the form: 
 \begin{equation}
 \label{eq1}
\dot{x}_k = \alpha f(x_k) +(1-\alpha) \left[ k \sum_{k'}P(k' \mid k) \frac{x_{k'}}{k'} - x_k \right] 
\end{equation}
where ${x}_k$ stands for the density displayed by the nodes that share the connectivity $k$. The discrete index $k$ runs from $1$ to $k_{max}$, where $k_{max}$ stands for the largest connectivity, as exhibited by the network being analyzed. $P(k' \mid k)$ is the conditional probability that a link exists from a given class $k$ to a class $k'$. In  (\ref{eq1}) we have implicitly assumed that the nonlinear contribution $f(x)$, can be also organized in classes $f(x_k)$, as reflecting the degree of connectivity associated to individual nodes. While this is not true in general, it can be reasonably postulated as long as $\alpha$ is forced small, i.e. when the system under scrutiny is a perturbation to the linear random walk problem.   Neglecting correlation among node degrees, one can break the probability as 
$P(k' \mid k) = \frac{k'P(k')}{\langle k \rangle}$, where $\langle k \rangle=\sum_{k} k P(k)$ and $P(k')$ is the connectivity distribution. This latter condition constitutes the core of the celebrated Heterogeneous Mean Field (HMF) approximation, to which we shall make extensive reference in the following.  A straightforward manipulation yields
\begin{equation}
\dot{x}_k = \alpha f(x_k) + (1-\alpha) \left[ \frac{k}{\langle k \rangle} \sum_{k'}P(k')x_{k'} - x_k \right] 
\end{equation}
Introduce now the quantity  $\Theta= \frac{1}{\langle k \rangle} \sum_{k'}P(k')x_{k'} $ which enables one to recast the previous equation in the compact form: 
\begin{equation}
 \label{eq2}
\dot{x}_k = \alpha f(x_k) + (1-\alpha)[k\Theta - x_k] 
\end{equation}
$\Theta$ is a collective mean-field variable, which allows to formally decouple the dynamics, as seen on different nodes, grouped in classes of homologous connectivity. Stated differently, the knowledge of $\Theta$ is sufficient, under the range of validity of the HMF approximation,  to solve for the densities at any time and for all degree classes $k$. In the following, we will focus on the equilibrium solution, which in turn amounts to setting $\dot{x}_k=0$ $\forall k$. We will then label with $\bar{x}_k$ the fixed points as displayed by the system and, consequently, $\bar{\Theta} =  \frac{1}{\langle k \rangle} \sum_{k'}P(k')\bar{x}_{k'}$.
Further, we will assume the nonlinear function $f(\cdot)$ to be of the logistic type, and thus set $f(\bar{x}_{k})=\bar{x}_{k}(1-\bar{x}_{k})$. This is not a mandatory step for the forthcoming analysis, as any generic nonlinear function would serve equally well the scope. The advantage of using a logistic equation resides in that it allows for explicit analytical progress to be made.

\begin{figure}[!htb]
\includegraphics[scale=0.35]{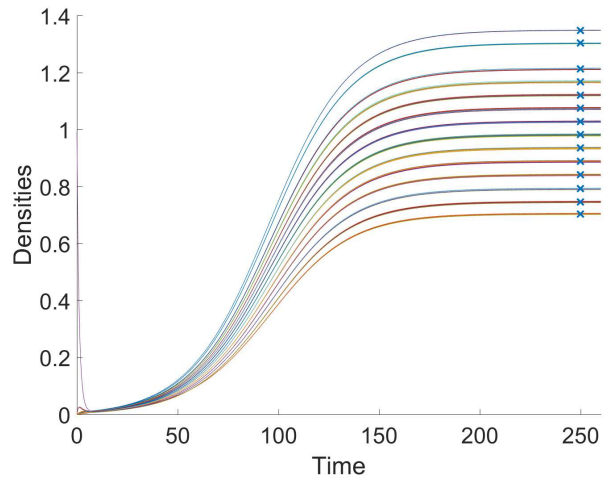}
\centering
\caption{Densities $x_i$ are plotted against time, starting out-of-equilibrium and assuming a random Watts-Strogatz network with relocation probability $\beta=0.99$
and  $N=200$ nodes. Solid lines refer to a direct integration of the governing Eqs.  (\ref{eq0}), while symbols (crosses) stand for the HMF-based prediction, obtained following the procedure described in the main text. The densities cluster in families of homologous connectivity $k$. Here, $\alpha=0.05$.}
\label{ForwardPreditionsExperiments}
\end{figure}

From Eq. (\ref{eq2}), at the fixed point, one gets:
\begin{equation}
\label{eq3}
\bar{x}_k=\frac{(2\alpha - 1)\pm \sqrt{(2\alpha - 1)^2 + 4\alpha[(1-\alpha)k\bar{\Theta} ]}}{2\alpha} 
\end{equation}
To elaborate on the fundamental interest of  Eq.  (\ref{eq3}), we consider a numerical implementation of system (\ref{eq0}), assuming a random network made of $N=200$ nodes (see caption fo Fig. \ref{ForwardPreditionsExperiments}) 
as the backbone support. Starting out of equilibrium, the system evolves towards a fixed point, as it can be appreciated by visual inspection of Fig. \ref{ForwardPreditionsExperiments}. Trajectories stemming from nodes sharing the same connectivity cluster together, thus confirming a posteriori the validity of the HMF ansatz. The asymptotic attractor as attained by the system in its late time evolution can be effectively estimated by resorting to  
relation (\ref{eq3}). More specifically, we select a randomly chosen node of the pool, with degree $k^*$, and measure the density therein displayed, $\bar{x}_{k^*}$.  By inversion of (\ref{eq3}) one gets an estimate for the mean-field variable $\bar{\Theta}$ as:
\begin{equation}
\bar{\Theta}=\frac{\alpha \bar{x}_{k^*}^2 - (2\alpha - 1)\bar{x}_{k^*} }{(1-\alpha)k^*}
\end{equation}
This latter is then inserted in equation (\ref{eq3}) to predict the equilibrium solution $\bar{x}_k$ for all choices of the class index $k$. The predicted values are depicted in 
Fig. \ref{ForwardPreditionsExperiments} with a symbol (crosses) and match the equilibrium solution as obtained by direct integration of the dynamics. This observation forms the basis of the scheme of inversion that we shall outline in the following. We remind that the inverse scheme is ultimately targeted to reconstructing the distribution of connectivity of a network, a priori unknown, that happens to host the inspected dynamics. Moreover, the number of necessary information are to be gathered on just one node.

\section{The inverse protocol}

The procedure that we shall illustrate builds on the following recipe. Imagine to perform a series of experiments by tuning progressively the parameter $\alpha$, in discrete, ascending steps. 
The sequence of the experiments is indexed by $r$, which ranges from $1$ to at least $k_{max}$. In each experiments the system is let to equilibrate, and the corresponding density $\bar{x}_{k^*}^{(r)}$ is recorded on a node of class $k^*$ where the inspection is performed. From the knowledge of  $\bar{x}_{k^*}^{(r)}$ one can infer an estimate of $\bar{\Theta}_{r}$, which can be used to access an approximate measure of $\bar{x}_{k}^{(r)}$, for $k \ne k^*$, by means of Eq. (\ref{eq3}). Combining these information together, and recalling the definition of $\Theta$, results in a linear problem for the unknown entries of the $q$-component vector 
$\vec{P}=(P(1) \hdots P(q))$. In formulae:

\begin{equation}
\langle k \rangle
\begin{bmatrix}
\bar{\Theta}_1 \\ \vdots \\ \bar{\Theta}_q
\end{bmatrix}
=
\underbrace{\begin{bmatrix}
\bar{x}^{(1)}_{1} & \cdots & \bar{x}^{(1)}_{q}\\
\vdots         & \ddots & \vdots\\
\bar{x}^{(q)}_{1}& \cdots  & \bar{x}^{(q)}_q
\end{bmatrix}}_{\doteq  \mathbf{\Gamma}}
\begin{bmatrix}
P(1)\\
\vdots\\
P(q)
\end{bmatrix}
\label{linearProb}
\end{equation}

Solving the above problem for $(P(1) \hdots P(q))$ implies inverting the matrix $\mathbf{\Gamma}$, a task that proved numerically cumbersome, being $\mathbf{\Gamma}$ poorly conditioned. To overcome this limitation we resorted to an optimization approach, which enforced the minimization of the norm $\| \vec{\Theta} \langle k \rangle - \mathbf{\Gamma}  \vec{P} \|$, while imposing the entries of  $\vec{P}$ to be positive defined. Here, $\vec{\Theta}= (\Theta_1  \hdots  \Theta_q)$. The average connectivity  $\langle k \rangle$ is a priori unknown and it is therefore assumed, as a free control parameter in the optimization scheme. More specifically, we set $\langle k \rangle$ to a nominal value and run consequently the optimization protocol, recording as an output the quantity $\sum_{k'} P(k')$. An implicit requirement of the analysis that leads to (\ref{eq3}) is the normalization of the distribution of connectivity,  $\sum_{k'} P(k')=1$. Among the solutions that are found by solving the problem in norm for different $\langle k \rangle$, we select the one that minimizes the positive residue $\left( 1- \sum_{k} P(k) \right)^2$. By invoking this closure of the scheme, we also get an estimate for the average connectivity $\langle k \rangle$. This latter could be challenged against the true values in synthetic network model, in the aim of testing the adequacy of the proposed procedure \footnote{In principle, one could absorb  $\langle k \rangle$ in the definition of $P_k$, compute the rescaled entries $\tilde{P}_k = P_k/\langle k \rangle$ via the linear problem and enforce  a posteriori the normalization. This scheme proved however less stable that the one that we have illustrated in the main body of the paper.} .

\begin{figure}
\includegraphics[scale=0.35]{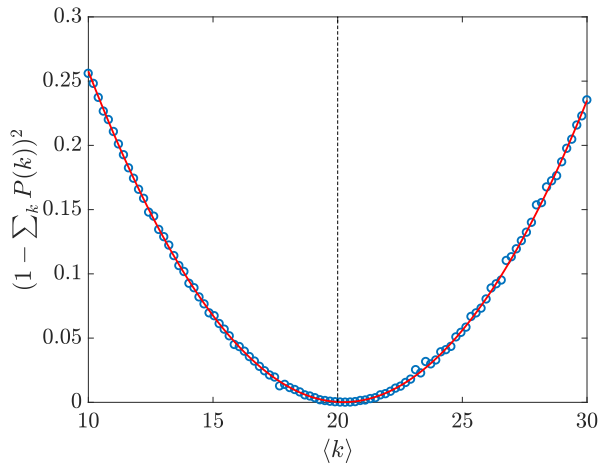}
\centering
\caption{The normalization error $\left( 1- \sum_{k} P(k) \right)^2$ is plotted against the value of $\langle k \rangle$ imposed when running the inverse scheme. A clear minimum is displayed for  
$\langle k \rangle \simeq 20$, which is very close to the correct value of the average connectivity. The reconstruction procedure is hence able to single out the correct average connectivity as possessed by the network being analyzed. The solid line is obtained by fitting a parabola to the collected data. }
\label{SumPKError}
\end{figure}

To this end we begin by considering the model (\ref{eq0}) defined on a random network made of $N=200$ nodes. The network is generated with the Watts-Strogatz recipe \cite{WS}, for a large relocation probability, which makes the network 
completely random. The average connectivity is $\langle k \rangle=20$. We performed $q=75$ different measurements, sampling the dynamics on the very same node and letting $\alpha$ to change in uniform steps in the interval $[0.005,0.4]$. In Fig. \ref{SumPKError} the normalization error $\left( 1- \sum_{k} P(k) \right)^2$ is depicted against the imposed average connectivity. A clear minimum is displayed, for approximately the correct value of $\langle k \rangle$.

\begin{figure}[!htb]
\includegraphics[scale=0.35]{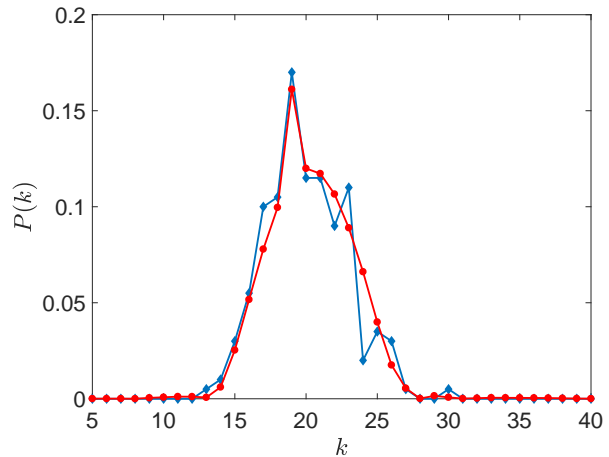}
\centering
\caption{The reconstructed distribution of connectivities: the blue line (with the diamond markers) represent the true degree distribution. The red line (with dot markers) stands for the distribution reconstructed via the procedure described in the main text. Here, $75$ independent experiments are employed and the dynamics is sampled from just one node of the collection. The network is generated according to the Watts-Strogatz recipe with relocation probability $\beta=0.99$. Here, $N=200$}
\label{BestAlphaWWS}
\end{figure}

\begin{figure}[!htb]
\includegraphics[scale=0.35]{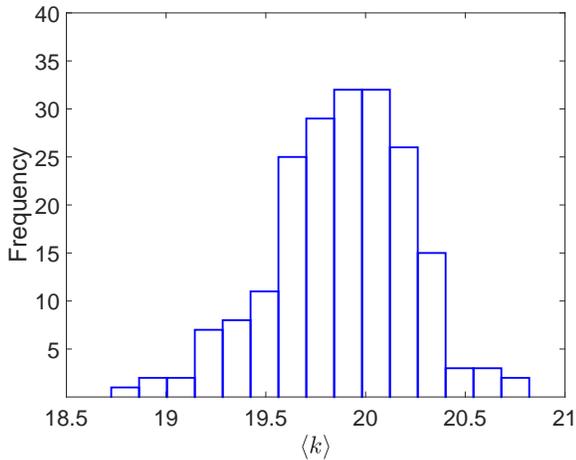}
\centering
\caption{The histogram of the averaged connectivity $\langle k \rangle$ is displayed. Each value of $\langle k \rangle$  refers to a different selection of the node from which the dynamics is sampled.}
\label{HistEK}
\end{figure}

\begin{figure}[!htb]
\includegraphics[scale=0.4]{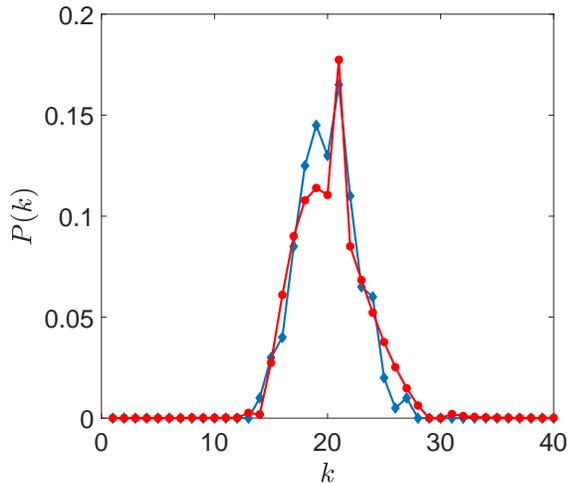}
\centering
\caption{The reconstructed distribution of connectivities, for a  Watts-Strogatz network with relocation probability $\beta=0.5$ and $N=200$. Symbols are chosen as explained in the caption of Fig. \ref{BestAlphaWWS}.}
\label{BestAlphaWWS1}
\end{figure}

Setting $\langle k \rangle$ to the value that minimizes the normalization error returns the distribution of connectivity depicted in Fig. \ref{BestAlphaWWS}. The blue line (with diamonds) stands for 
the true distribution, while the red curve (with dot markers) refer to the reconstructed profile. Changing the node from which the dynamics is sampled yields different estimates of the average connectivity $\langle k \rangle$ (and of the distribution that is consequently recovered). To provide a qualitative illustration of the degree of variability that stems from an arbitrary choice of the reference node, we plot in Fig. 
\ref{HistEK} the histogram of $\langle k \rangle$, obtained for all possible selections of the observation site. The distribution of predicted average connectivity is peaked around the correct solution. To improve on the accuracy of the method one can repeat the measurements on different sites and combine together the acquired information. This significantly improve on the ability of the HMF approximation to
adhere on the exact asymptotic solution, as seen in direct simulations. In Fig. \ref{BestAlphaWWS1} the reconstruction procedure is tested for a Watts Strogatz network with a smaller relocation probability and the quality of the reconstruction is still satisfying.

\begin{figure}[!htb]
\includegraphics[scale=0.4]{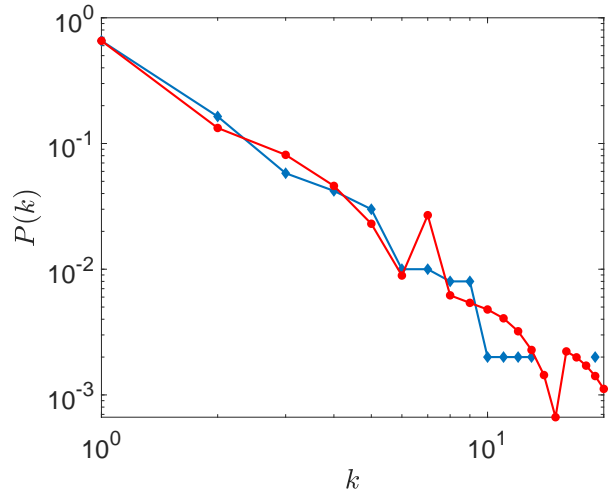}
\centering
\caption{The reconstructed distribution, assuming a scale free network. The network consists of $500$ nodes. Here, $P(k) \propto k^{-\gamma}$ with $\gamma=3$. Symbols follows the convention introduced in Fig. \ref{BestAlphaWWS}
}
\label{AlphaWBestScaleFree}
\end{figure}

To further elaborate on the potential of the described procedure we considered a scale free network made of $N=500$ nodes. More specifically $P(k) \propto k^{-\gamma}$ with $\gamma=3$. The analysis returns 
$\langle k \rangle= 2.01$ in excellent agreement with the correct value and the recovered $P(k)$ is displayed in Fig. \ref{AlphaWBestScaleFree}. At variance to the case of the random network analyzed above, changing the reference node impacts more significantly on the recorded $P(k)$ (while $\langle k \rangle$ is always correctly estimated). Averaging independent profiles obtained by sampling the dynamics from distinct nodes contributed to enhance the fidelity of the reconstruction .

Summing up to this point, we have introduced and successfully applied a procedure to reconstruct the distribution of connectivity of an unknown network by performing local measurement on just on node. The inspected system combines nonlinear reactions to relocation.  The idea of modulating their relative balance via the scalar parameter $\alpha$, might sound however artificial and with a reduced applied interest. 
To overcome this intrinsic limitation we shall consider in the following a variant of the dynamical model.

\section{Introducing the sources and modulating their strengths}

The updated scheme builds on the following steps. First, we freeze $\alpha$ to a constant amount. Then, we modify the dynamics of a given class of nodes, say those characterized by a connectivity $\bar{k}$, by introducing an ensemble made of identical sources, characterized by a constant injection strength $\eta$. Mathematically,  this corresponds to inserting on the right hand side of Eq. (\ref{eq0}) a constant factor $\eta$, for the entries $i$ which identify the selected pool of nodes. For any given choice of $\eta$, the asymptotic fate of the system can be analytically investigated by proceeding with the HMF approximation, in analogy with the above. Formally, one ends up with an expression for the fixed point which trivially extends that displayed in Eq. (\ref{eq3}):

\begin{equation}
\label{eq4}
\bar{x}_k=\frac{(2\alpha - 1)\pm \sqrt{(2\alpha - 1)^2 + 4\alpha[(1-\alpha)k\bar{\Theta} + \eta \delta_{k \bar{k}}}]}{2\alpha} 
\end{equation}

where $\delta_{k \bar{k}}$ stands for the Kronecker delta.

\begin{figure}[!htb]
\includegraphics[scale=0.38]{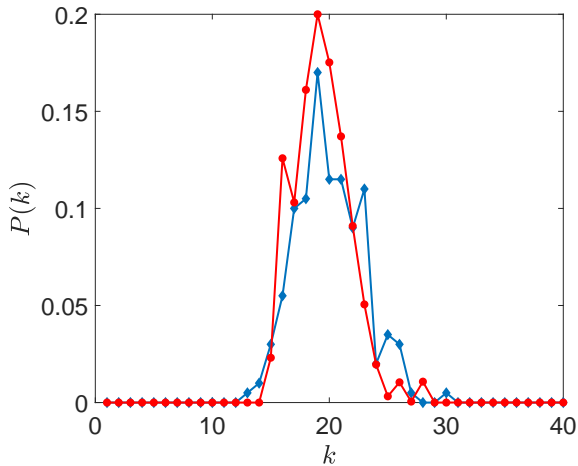}
\centering
\caption{The reconstructed distribution of connectivities: the blue line (with the diamond markers) stands for the true degree distribution. The red line (with dot markers) identifies the distribution reconstructed via the procedure described in the main text. Here, $\alpha=0.005$ and $\eta$ is changed in the interval $[0.005, 5.5]$ with uniform increments of $0.2$. The sources are placed on the nodes sharing degree $\bar{k}=15$ (the results do not depend on this specific choice). This class contains 7 different nodes, for the network realization here considered. The network employed is that of Fig.
\ref{BestAlphaWWS}.}
\label{BestEta}
\end{figure}

The reconstruction scheme can be hence modified as follows: (i) change $\eta$ within a given interval; (ii) for each choice of  $\eta$, measure the asymptotic state attained by the system on a given node with degree $k=k^*$; (iii) use this knowledge to estimate the mean-field variable $\bar{\Theta}$, by inversion of Eq. (\ref{eq4}) and, finally, (iv) predict the equilibrium solution $\bar{x}_k$, $\forall k \ne k^*$. Repeating this procedure for a sufficiently large set of distinct values of $\eta$, yields a linear problem of the type (\ref{linearProb}) which can be solved in norm to compute $\vec{P}=(P(1) \hdots P(q))$. Also in this case the average connectivity is estimated by minimizing the residual error.  As a demonstrative example, we compare in Fig. \ref{BestEta} the reconstructed $P(k)$ to its exact homologue. Here, the underlying network is generated according to the Watts-Strogatz recipe and the agreement between the two depicted curves is satisfying.  

As a final point, we will relax the assumption of dealing with a full class of nodes which behave as identical sources. More precisely, we will break the symmetry and assume that just one node of a 
given class $\bar{k}$ acts as a source. To handle this generalized setting, we revisit the definition of  the mean field variable. Recall that $P(\bar{k})$ measures the number of nodes sharing connectivity  
$\bar{k}$ normalized to the system size $N$. Then one can define the following collective variable:

\begin{equation}
\label{newglobal}
\Pi= \frac{1}{\langle k \rangle} \left[ \sum_{k'}P(k')x_{k'} + \frac{y -x_{\bar{k}}}{N} \right]
\end{equation}

where $y$ denotes the value of the density at the source location. One can apply to HMF machinery\footnote{In doing so we postulate that the inserted source does not disrupt the organization in classes. The validity of this working ansatz is confirmed a posteriori by the quality of the obtained reconstruction.} to yield:

\begin{equation}
\label{N1}
\bar{x}_k=\frac{(2\alpha - 1)\pm \sqrt{(2\alpha - 1)^2 + 4\alpha(1-\alpha)k \Pi}}{2\alpha}
\end{equation}

for the fixed point concentration as predicted on all nodes, but the source. This latter is characterized by an asymptotic density given by:

\begin{equation}
\label{N2}
\bar{y}=\frac{(2\alpha - 1)\pm \sqrt{(2\alpha - 1)^2 + 4\alpha[(1-\alpha)k \Pi + \eta]}}{2\alpha}
\end{equation}

\begin{figure}[!htb]
\includegraphics[scale=0.4]{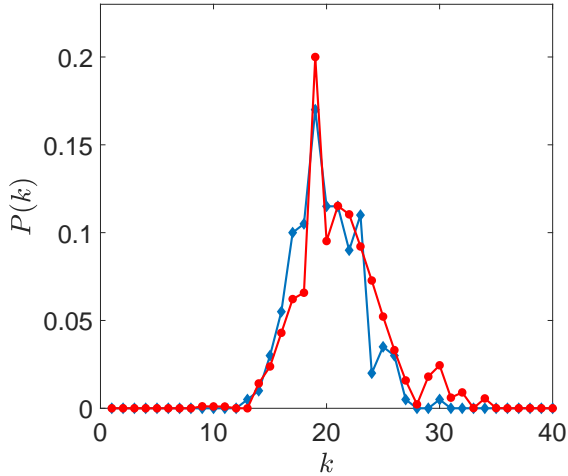}
\centering
\caption{The reconstructed distribution of connectivities: the blue line (with the diamond markers) stands for the true degree distribution. The red line (with dot markers) identifies the distribution reconstructed via the procedure described in the main text. Here, $\alpha=0.005$ and $\eta$ is changed in the interval $[0.005, 0.35]$, with $80$ successive uniform increments. The source is placed on one of the nodes sharing degree $\bar{k}=15$. The network employed is that of Fig.
\ref{BestAlphaWWS}.}
\label{OneWSBest}
\end{figure}

Building on the above, one can put forward a straightforward generalization of the reconstruction algorithm. For any given choice of the source injection rate $\eta$, one can measure the equilibrium density as displayed on one individual node, belonging to class $k^*$. By manipulating Eq. (\ref{N1}), one can then estimate $\Pi$ and use this latter to predict the expected density on each of the classes ($\bar{x}_k$, $k \ne k^*$) and on the source ($\bar{y}$). To this end one makes explicit use of, respectively, Eqs. (\ref{N1}) and (\ref{N2}). Repeating the analysis for a sufficiently large set of  $\eta$ returns a linear problem of the type discussed above, with the sole difference that now the size of the network $N$ qualifies as one of the unknowns to be eventually recovered. In formulae, Eq. (\ref{newglobal}) can be cast in the form:

\begin{equation*}
\langle k \rangle \Pi=  \sum_{k'}P(k')x_{k'} + \frac{z}{N}
\end{equation*}

where $z=y -x_{\bar{k}}$. We hence get:

\begin{equation}
\langle k \rangle
\begin{bmatrix}
\bar{\Pi}_1 \\ \vdots \\ \bar{\Pi}_{q}
\end{bmatrix}
=
\begin{bmatrix}
\bar{x}^{(1)}_{k_1} & \cdots & \bar{x}^{(1)}_{q} & \bar{z}_1\\
\vdots         & \ddots & \vdots & \vdots\\
\bar{x}^{(q)}_{1}& \cdots  & \bar{x}^{(q)}_{q} & \bar{z}_{q}
\end{bmatrix}
\begin{bmatrix}
P(1)\\
\vdots\\
P(q)\\
\frac{1}{N}
\end{bmatrix}
\end{equation}

where $q$ ($> k_{max}$) stands for the number of repeated measurements performed at different choices of $\eta$. The reconstruction obtained following this updated strategy is displayed in Fig. 
(\ref{OneWSBest}), for the same realization of the Watts-Strogatz network as considered above. The quality of the reconstruction is still adequate and the estimated value of $N$ is in agreement with the true one. 
The performance of the algorithm may change depending on the node of observation but the distribution of $N$ obtained when covering the full set of possible choices is peaked in correspondence of the correct value. Similarly the average value of  $\langle k \rangle$ is found to be correctly estimated. 

\section{Conclusion}

In conclusion, we have here introduced and tested a procedure to access structural information on network topology. The method samples the dynamics of reactive walkers, microscopic 
entities which are made to explore the embedding network, while subject to nonlinear (and local) reaction terms. The imposed non-linearity makes it possible to recover the network's distribution of connectivity, from sequences of measurements performed on just one node of the collection. To reach this goal, we exploit the organization in classes of the ensuing dynamical equilibrium and make explicit use of the celebrated Heterogeneous Mean Field approximation. A variant of the method which consists in introducing localized sources of modulable strength, enables in turn to estimate the size of the scrutinized network.
Future investigations will be aimed to improving on the optimization scheme and, consequently, on testing the predictive ability of the proposed techniques versus more challenging network architecture, 
including multiplex.

\end{document}